\documentclass[journal]{IEEEtran}

\ifCLASSINFOpdf
\else
   \usepackage[dvips]{graphicx}
\fi
\usepackage{url}

\usepackage{graphicx}

\usepackage[T1]{fontenc}
\usepackage[latin9]{inputenc}
\usepackage{pgf}
\usepackage{xcolor}
\usepackage{physics}
\usepackage{notation}
\usepackage{multirow}
\usepackage{amssymb}
\usepackage{acronym}
\usepackage{psfrag}
\usepackage{soul}
\usepackage{tikz}
\usepackage{subfig}
\usepackage{cite}
\usepackage[inline]{enumitem}
\usepackage[level=3]{LatexInclusion/messages}
\acrodef{pmf}[PMF]{probability mass function}
\acrodef{pdf}[PDF]{probability density function}
\acrodefplural{pdf}[PDFs]{probability density functions}
\acrodef{mmse}[MMSE]{minimum mean-square error}
\acrodef{da}[DA]{data association}
\acrodef{bp}[BP]{belief propagation}
\acrodef{fg}[FG]{Factor Graph}
\acrodef{nebp}[NEBP]{neural enhanced belief propagation}
\acrodef{roi}[ROI]{region of interest}
\acrodef{bev}[BEV]{bird's eye view}
\acrodef{mot}[MOT]{multi-object tracking}
\acrodef{po}[PO]{potential object}
\acrodef{jpda}[JPDA]{joint probabilistic data association}
\acrodef{mht}[MHT]{multiple hypothesis tracker}
\acrodef{rfs}[RFS]{random finite sets}
\acrodef{gospa}[GOSPA]{generalized optimal sub-pattern assignment}
\acrodef{mospa}[MOSPA]{mean optimal sub-pattern assignment}
\acrodef{tbd}[TBD]{track-before-detect}
\acrodef{mb}[MB]{multi-Bernoulli}
\acrodef{iemb}[IEMB]{information exchange multi-Bernoulli}
\acrodef{snr}[SNR]{signal-to-noise ratio}
\acrodef{iid}[IID]{independent and identically distributed}

\newcommand{\ist}{\hspace*{.3mm}}
\newcommand{\rmv}{\hspace*{-.3mm}}
\newcommand{\iist}{\hspace*{1mm}}
\newcommand{\rrmv}{\hspace*{-1mm}}
\newcommand{\nn}{\nonumber}
\newcommand{\T}{\mathrm{T}}

\begin{document}

\title{A BP Method for Track-Before-Detect}

\author{Mingchao~Liang,~\IEEEmembership{Student Member,~IEEE,}
Thomas~Kropfreiter,~\IEEEmembership{Member,~IEEE,}
        and~Florian~Meyer,~\IEEEmembership{Member,~IEEE}

\thanks{This work was supported in part by the National Science Foundation (NSF) under CAREER Award No. 2146261.}

\thanks{Mingchao~Liang, Thomas~Kropfreiter, and Florian~Meyer are with the  University of California San Diego, La Jolla, CA, USA (e-mail: \texttt{m3liang@ucsd.edu, tkropfreiter@ucsd.edu, flmeyer} \texttt{@ucsd.edu,}).}
\vspace{-7.5mm}
}

\markboth{IEEE Signal Processing Letters, 2023}
{Shell \MakeLowercase{\textit{et al.}}: Bare Demo of IEEEtran.cls for IEEE Journals}
\maketitle

\begin{abstract}
Tracking an unknown number of low-observable objects is notoriously challenging. 
This letter proposes a sequential Bayesian estimation method based on the \ac{tbd} approach. 
In \ac{tbd}, raw sensor measurements are directly used by the tracking algorithm without any preprocessing. 
Our proposed method is based on a new statistical model that introduces a new object hypothesis for each data cell of the raw sensor measurements. It allows objects to interact and contribute to more than one data cell. 
Based on the factor graph representing our statistical model, we derive the message passing equations of the proposed \ac{bp} method for \ac{tbd}. 
Approximations are applied to certain \ac{bp} messages to reduce computational complexity and improve scalability.
In a simulation experiment, our proposed \ac{bp}-based \ac{tbd} method outperforms two other state-of-the-art \ac{tbd} methods.
\end{abstract}

\begin{IEEEkeywords}
Multi-object tracking, track-before-detect, factor graph, belief propagation.
\vspace{-3mm}
\end{IEEEkeywords}

\IEEEpeerreviewmaketitle

%-------------------------------------------------------------------------------------

%-------------------------------------------------------------------------------------
\section{Introduction}
\label{sec:Intro}
Multi-object tracking \acused{mot}(\ac{mot}) \cite{BarWilTia:B11,Mah:B07,ChaMor:B11,MeyBraWilHla:J17,MeyKroWilLauHlaBraWin:J18,Wil:J15} aims at estimating the number and states of a time-varying number of objects from noisy sensor measurements. 
Possible applications of \ac{mot} include applied ocean science \cite{FerMunTesBraMeyPelPetAlvStrLeP:J17}, indoor localization \cite{WitMeiLei:J16}, and autonomous driving \cite{MeyWil:J21}.
In the conventional \textit{detect-then-track} approach, a detection stage preprocesses the raw sensor data in order to reduce data flow and computational complexity. 
The resulting ``point measurements'' are then the input to the tracking stage.
However, this preprocessing leads to a loss of relevant information and thus to a reduced tracking performance, especially in low \ac{snr} scenarios. 
%In this way, only a reduced number of data cells are considered. The detection stage reduces data flow and computational complexity but can lead to a loss of relevant information and, especially in scenarios with low \ac{snr}, a reduced tracking performance.
%However, this preprocessing step introduces spurious non-object generated measurements and a data association uncertainty, which means that now it is unknown which measurement was generated by which object. 

 %assumption that an object can be associated with at most one measurement, and vice versa. 

%Conventional methods formulate and solve the \ac{mot} problem in the Bayesian estimation framework \cite{BarWilTia:B11,Mah:B07,MeyBraWilHla:J17,MeyKroWilLauHlaBraWin:J18,Wil:J15,GarWilGraSve:J18} and have good performance. However, these conventional \ac{mot} methods make the data association assumption that an object can be associated with at most one measurement, and vice versa. 
%Hence, they rely on a detector to preprocess the raw sensor data and to provide a significantly smaller number of measurements for downstream tracking algorithms. 
%This process is referred to as \textit{detect-then-track}. On one hand, detect-then-track reduces time complexity. On the other hand, the preprocessing step may lose some critical tracking information, especially for weak objects that lead to low \ac{snr} measurements. 

In \textit{Track-before-detect} \acused{tbd}(\ac{tbd}) methods, raw sensor data is passed directly to the tracking stage without any preprocessing.
 %and the resulting information loss.  
State-of-the-art \ac{tbd} methods may be distinguished between batch processing approaches and sequential Bayesian estimation methods.
Batch processing approaches include methods based on maximum likelihood estimation \cite{TonBar:J98}, the Hough transform \cite{MoySpaLam:J11}, and dynamic programming \cite{Bar:J85}.
 %were used proposed for \ac{tbd}, but they have high computational complexity.
On the other hand, sequential Bayesian estimation methods 
%\cite{TonBar:J98,MoySpaLam:J11,Bar:J85,DavGae:B18,RisVoVoFar:J13,RisRosKimWanWil:J20,VoVoPhaSut:10,KimRisGuaRos:21,DavGar:22,KroWilFlo:21,CaoZha:J22} 
perform estimation of object states represented either by random vectors \cite{SalBir:01,OrtFit:J02,BoeDriVerHeeJul:03,ItoGod:J13,DavGae:B18} or \ac{rfs} \cite{RisVoVoFar:J13,RisRosKimWanWil:J20,VoVoPhaSut:10,KroWilFlo:21,KimRisGuaRos:21,DavGar:22,CaoZha:J22}.
While vector-based approaches often rely on particle filtering methods \cite{SalBir:01,OrtFit:J02,BoeDriVerHeeJul:03,ItoGod:J13}, set-based methods are mostly based on the Bernoulli filter for single object tracking \cite{RisVoVoFar:J13,RisRosKimWanWil:J20} or its generalizations for multi-object tracking \cite{VoVoPhaSut:10,KroWilFlo:21,KimRisGuaRos:21,DavGar:22,CaoZha:J22}.  
%\cite{RisVoVoFar:J13,RisRosKimWanWil:J20} solved \ac{tbd} using particle filter, but they only consider the single object case.}
 Recently introduced \ac{tbd} methods that are suitable for tracking an unknown number of objects (i) assume non-interacting objects, i.e., regions of measurements influenced by different objects do not overlap \cite{VoVoPhaSut:10}; (ii) rely on heuristics to introduce newborn objects\cite{KimRisGuaRos:21,DavGar:22,CaoZha:J22}; or (iii) assume that every object can contribute to at most one measurement \cite{KroWilFlo:21}.
%However, none of these methods have a statistical birth model, i.e., they rely on certain heuristics \cite{KimRisGuaRos:21}, or even the true start position of objects \cite{VoVoPhaSut:10,DavGar:22}, to initialize objects  that appear in the scene for the first time.
%The goal of this paper is to develop an \ac{tbd} method for \ac{mot} that can address the aforementioned limitations.

Belief propagation (\ac{bp}) \cite{KscFreLoe:01,YedFreWei:05,KolFri:B09,Loe:04}, also known as the sum-product algorithm, is a versatile and efficient method for performing inference in Bayesian estimation problems.
More precisely, \ac{bp} has already been used very successfully in detect-then-track \ac{mot} problems \cite{MeyBraWilHla:J17,MeyKroWilLauHlaBraWin:J18}. 
Here, the statistical model underlying the \ac{mot} problem is represented by a so-called factor graph.
By computing local messages and sending them along the edges through the graph, the structure of the \ac{mot} model can be exploited in order to reduce computational complexity and increase scalability \cite{MeyBraWilHla:J17,MeyKroWilLauHlaBraWin:J18}.   

%the statistical model of the \ac{mot} problem is represented by a so-called factor graph, and \ac{bp} performs local operations called ``messages'' on the edges of the graph. This makes it possible to exploit the structure of the statistical model to reduce computational complexity and increase scalability \cite{MeyBraWilHla:J17,MeyKroWilLauHlaBraWin:J18}.
%\ac{mot} factor graph, the \ac{bp}-based \ac{mot} approach turns out to be both powerful and highly scalable 
%, which can maintain a large number of object tracks and provide state-of-the-art tracking performance. 
%\rd{CITE FUSION, more pixels affected}

%The goal of this paper is to develop an \ac{tbd} method for \ac{mot} that can address the aforementioned limitations. 
%\ac{bp}, also known as the sum-product algorithm, \cite{KscFreLoe:01,YedFreWei:05,KolFri:B09} is a good candidate as it has been successfully applied to conventional \ac{mot} problems \cite{MeyBraWilHla:J17,MeyKroWilLauHlaBraWin:J18}. The \ac{bp}-based \ac{mot} methods \cite{MeyBraWilHla:J17,MeyKroWilLauHlaBraWin:J18} represent the statistical model using a factor graph, and run the \ac{bp} algorithm by ``passing messages'' along the edges of the factor graph. By exploiting the structure of the factor graph, \ac{bp}-based \ac{mot} methods are highly scalable, which can maintain a large number of object tracks and provide state-of-the-art tracking performance. \rd{CITE FUSION, more pixels affected}

In this letter, we propose a \ac{bp} method for \ac{tbd}.
More precisely, we propose a new statistical model for the \ac{tbd} problem,
develop the corresponding factor graph, and perform inference by applying BP on that graph.
%Within our Bayesian estimation framework, object states and measurements are modeled by random vectors.
%We model object states and measurements 
%In our approach, Since \acp{po} and data cells are inherently ordered, we use random vectors to represent object states and measurements within our Bayesian estimation framework.
Our statistical model includes a new measurement model in which interacting objects can contribute to more than one data cell.
In fact, this new measurement model can be considered a generalization of other models used in existing TBD methods.
Furthermore, a new object hypothesis, referred to as \ac{po}, is introduced for every cell measurement. 
%Since \acp{po} and data cells are inherently ordered, we use random vectors to represent object states and measurements within our Bayesian estimation framework. 
%We furthermore propose a new measurement model that considers the interaction between between objects and models
 %that generalizes most models used in the literature, allows interacting objects to contribute to more than one data cell. 
%We also present the factor graph for the new statistical model and derive the corresponding \ac{bp} messages. 
To reduce computational complexity, certain BP messages are approximated by Gaussian \acp{pdf} using moment matching. 
This approximation is similar to the one performed within the \ac{rfs}-based \ac{tbd} method with heuristic track initialization in \cite{DavGar:22}.  
The main contributions of this letter can be summarized as follows\vspace{.5mm}:
\begin{itemize}
    \item We propose a new statistical model for \ac{tbd} \ac{mot} consisting of a new measurement model for interacting objects and a new model for object birth.
    \vspace{1.5mm}
    \item We derive a scalable \ac{bp} inference method and demonstrate its improved performance compared to two other state-of-the-art \ac{tbd} algorithms.
    \vspace{.5mm}
\end{itemize}
To the best of the author's knowledge, the presented approach is the first \ac{tbd} method based on \ac{bp}.
\vspace{-.5mm}
%-------------------------------------------------------------------------------------

%-------------------------------------------------------------------------------------
\section{System Model} \label{sec:sys_model}
\vspace{-.5mm}

%The number of objects in the considered \ac{tbd} problem is time-varying and unknown. 
We model the multi-object state at discrete time $k$ by $N_k$ \acp{po} \cite{MeyBraWilHla:J17,MeyKroWilLauHlaBraWin:J18}, where the existence of each \ac{po} $n \in \{1, \dots, N_k\}$ is described by the binary random variable $r_{k, n} \in \{0,1\}$. Here, $r_{k, n} \rmv=\rmv 1$ indicates that \ac{po} $n$ exists. 
%$N_k$ is the maximum possible number of objects at time step $k$. 
The kinematic state of \ac{po} $n$ is modeled by the random vector $\V{x}_{k, n}$ whose entries describe the object's position, the object's intensity, and possibly further kinematic properties of the object. 
We define the state of \ac{po} $n$ by $\V{y}_{k, n} = [\V{x}_{k, n}^\T \hspace{1mm} r_{k, n}]^\T$ and the joint state of all \acp{po} by $\V{y}_{k} = [\V{y}_{k, 1}^\T \cdots \V{y}_{k, N_k}^\T]^\T$\rmv. 
For later use, we furthermore define $\V{r}_{k} = [r_{k, 1} \cdots r_{k, N_k}]^\T$.
\vspace{-1mm}

%To take objects that newly appear in the scene into account, one new \ac{po} is introduced for each of the $J$ measurements. Together with the $N_{k - 1}$ \acp{po} from previous time steps, the total number of \acp{po} at time $k$ is $N_k = N_{k - 1} + J$. In what follows, we use the first $N_{k - 1}$ indices $n \in \{1, \dots, N_{k - 1}\}$ to denote the \acp{po} that have been introduced at previous time steps, and the last $J$ indices $n \in \{N_{k - 1} + 1, \dots, N_{k - 1} + J\}$ to denote the \acp{po} that are newly introduced at\vspace{-3mm} time $k$.

\subsection{Superpositional Measurement Model}
\label{subsec:sys_model_meas}
\vspace{-.5mm}

Our measurement model consists of $J$ data cells, where each cell $j$ can be associated with a time-varying intensity measurement $\V{z}_{k, j}$.
We stack all measurements $\V{z}_{k, j}$ into the joint measurement vector $\V{z}_k = [\V{z}_{k, 1}^\T \cdots \V{z}_{k, J}^\T]^\T$.
 %measurements $\V{z}_k = [\V{z}_{k, 1}^\T \cdots \V{z}_{k, J}^\T]^\T$ at each time step that, in a TBD problem, typically represent the intensity of data cells, e.g., the pixels in an image. 
Note that the positional information of $\V{z}_{k, j}$ is encoded by its index $j$.
We now model $\V{z}_{k, j}$, according to
%Each data cell is modeled by the\vspace{-.2mm} vector
\vspace{-1mm}
\begin{equation}
    \V{z}_{k,j} = \sum_{n = 1}^{N_k}\ist r_{k, n}\iist \V{h}_{j,k,n} + \V{\epsilon}_{k, j}\ist.  \label{eq:meas_model}
    \vspace{0mm}
\end{equation}
Here, $\V{h}_{j,k, n} \rmv\in\rmv \mathbb{R}^d$ is the contribution of \ac{po} $n$ to measurement $\V{z}_{k,j}$ if \ac{po} $n$ exists, i.e., if $r_{k, n} \rmv=\rmv 1$.
We model $\V{h}_{j,k, n}$ by the Gaussian \ac{pdf} $f\big(\V{h}_{j, k, n} \ist | \ist \V{x}_{k, n} \big) = \Set{N} \big(\V{h}_{j,k, n}; \V{\mu}_j(\V{x}_{k, n}), \M{C}_j(\V{x}_{k, n}) \big) $ whose mean $\V{\mu}_j(\V{x}_{k, n})$ and covariance matrix $\M{C}_j(\V{x}_{k, n})$ define a point spread function \cite{VoVoPhaSut:10, RisRosKimWanWil:J20,KimRisGuaRos:21,KroWilFlo:21}. 
We assume that $\V{h}_{j,k, n}$ is statistically independent for all $k$, $j$, and $n$, and also independent of $\V{\epsilon}_{k, j}$. 
Note that for $r_{k, n}\rmv =\rmv 0$, \ac{po} $n$ does not exist and hence does not contribute to any measurement. 
Furthermore, the additive noise component $\V{\epsilon}_{k, j}$ is modeled as Gaussian with zero mean and covariance $\M{C}_{{\epsilon}}$. It is further assumed statistically independent across all $k$ and $j$. 

From measurement equation \eqref{eq:meas_model}, we can directly infer the conditional \acp{pdf} $f(\V{z}_{k, j} | \V{y}_{k})$ by using the fact that the sum of statistically independent Gaussian variables, i.e., in our case all the $\V{h}_{j,k, n}$ and $\V{\epsilon}_{k, j}$, is again a Gaussian variable.
Furthermore, since $\V{h}_{j,k, n}$ and $\V{\epsilon}_{k, j}$ are also conditionally independent for all $j$ given $\V{y}_{k}$, all the measurements $\V{z}_{k,j}$ are in turn conditionally independent given $\V{y}_{k}$.
This leads to the joint likelihood function given by $f(\V{z}_{k} | \V{y}_{k}) = \prod_{j = 1}^{J} \ist f(\V{z}_{k, j} | \V{y}_{k})$.
Note that for some types of objects, e.g., objects with plane surfaces, the assumption of independent measurements $\V{z}_{k,j}$ does not hold because the intensity values of neighboring data cells are usually correlated. 
However, most \ac{tbd} algorithms rely on this model assumption, which is required for efficient estimation \cite{RisRosKimWanWil:J20,KimRisGuaRos:21,KroWilFlo:21,VoVoPhaSut:10,DavGar:22}. 
 
%On the one hand, neglecting the measurement correlation might result in a tracking algorithm of lower tracking accuracy compared to an algorithm that considers correlated measurements but on the other hand is a necessary condition to result in fast inference. 
%
 %the measurements $\V{z}_{k,j}$ are potentially correlated since an object might can potentially illuminate more than a single data cell, the $\V{z}_{k,j}$ are statistically dependent. However,     Note that modeling $\V{z}_{k,j}$ as conditionally independent given $\V{y}_{k}$ is quite common. It usually results in   
%the latter independence assumption over all $j$, is a common model assumption in many  

%$p(\V{z}_{k, j} | \V{h}_{k,j}, \V{r}_{k})$, where we have introduced $\V{h}_{k,j} = [\V{h}_{k, j,1}^\T \cdots \V{h }_{k, j, N_k}^\T]^\T$. 
%
%We can furthermore obtain the likelihood function $p(\V{z}_{k, j} | \V{y}_{k})  \rmv= \int \ist p(\V{z}_{k, j} | \V{h}_{k,j}, \V{r}_k) \ist \prod^{N_k}_{n=1} \ist p(\V{h}_{k, j, n} | \V{x}_{k, n}) \ist \mathrm{d} \V{h}_{k, j}$ as well as the joint likelihood function $p(\V{z}_{k} | \V{y}_{k}) = \prod_{j = 1}^{J} \ist p(\V{z}_{k, j} | \V{y}_{k})$. 
Note that our superpositional model in \eqref{eq:meas_model} generalizes many \ac{tbd} measurement models in the literature. In particular, by setting $\M{C}_j(\V{x}_{k, n}) \rmv=\rmv \M{0}$, our model reduces to the model used in \cite{VoVoPhaSut:10,DavGar:22}, by setting $d = 2$, $\V{\mu}_j(\V{x}_{k, n}) = \V{0}$, and $\M{C}_{\V{\epsilon}} = \sigma_{\epsilon}^2 \M{I}_d$, it is equivalent to the Rayleigh model in \cite{KroWilFlo:21}, and by additionally assuming that $\sigma_{\epsilon}^2$ is Gamma distributed, it is equal to the model in \cite{RisRosKimWanWil:J20,KimRisGuaRos:21}.

%\rd{Our superpositional intensity model in \eqref{eq:meas_model} generalizes the models used in \cite{VoVoPhaSut:10,DavGar:22} which was applied to image data from radar, sonar, or sensor network.} It is reduced to the model in \cite{VoVoPhaSut:10,DavGar:22} by setting $\M{C}_j(\V{x}_{k, n}) = \M{0}$, i.e., by making $\V{h}_{k, j, n}$ a deterministic function of $\V{x}_{k, n}$. \rd{Models for maritime radar data in \cite{RisRosKimWanWil:J20,KimRisGuaRos:21} can also be adapted from the model in \eqref{eq:meas_model} by setting $d = 2$, $\V{\mu}_j(\V{x}_{k, n}) = \V{0}$, and $\M{C}_{\V{\epsilon}} = \sigma_{\epsilon}^2 \M{I}_d$ with gamma-distributed $\sigma_{\epsilon}^2$. If $\sigma_{\epsilon}^2$ is a constant, it represents Rayleigh-distributed measurements \cite{KroWilFlo:21}.}

% \st{Without loss of information, it can also represent Rayleigh-distributed measurements \cite{KroWilFlo:21}. In particular, by sampling a uniformly distributed phase for each Rayleigh-distributed measurement, computing the corresponding complex-valued measurement, and stacking its real and imaginary parts, the resulting 2-D measurement vector follows our measurement model in \eqref{eq:meas_model}.}
% i.e, $\V{\epsilon}_{k, j} \sim \Set{N}(\V{\epsilon}_{k, j};\V{0}, \M{C}_{\V{\epsilon}})$,

\subsection{State-Transition and Birth Model}
\label{sec:sta_trans}
%\vspace{-1mm}

It is assumed that the legacy \ac{po} states $\V{y}_{k - 1, n}$, $n \in \{1, \dots, N_{k - 1}\} $ evolve independently in time \cite{BarWilTia:B11}. Thus,
%and identically distributed following a first-order Markov process \cite{BarWilTia:B11}, 
the joint state transition function can be factored according to \cite[Sec.~VIII-C]{MeyKroWilLauHlaBraWin:J18}
 %of the joint \ac{po} state $\V{y}_{k - 1}$ at time $k - 1$, can be expressed as
% \begin{equation}
%     f(\underline{\V{y}}{}_{k} | \V{y}_{k - 1}) = \prod_{n = 1}^{N_{k - 1}} f(\V{y}_{k, n} | \V{y}_{k - 1, n}) \label{eq:dynamics}
% \end{equation}
$ f(\underline{\V{y}}{}_{k} | \V{y}_{k - 1}) = \prod_{n = 1}^{N_{k - 1}} f(\V{y}_{k, n} | \V{y}_{k - 1, n})$, where we have introduced
%. Here we have used the notation 
$\underline{\V{y}}{}_{k} = [\V{y}_{k, 1}^\T \cdots \V{y}_{k, N_{k - 1}}^\T]^\T$ 
To account for newly appearing objects, we introduce, at each time $k$, $J$ new \acp{po} with states $\V{y}_{k, n}$, $n \rmv\in \{N_{k - 1}+1,\ldots,N_{k}\}$, one new \ac{po} for each data cell $j$.
Thus, $N_k \rmv= N_{k - 1} \rmv+ J$. 
We define $\overline{\V{y}}_{k} = [\V{y}_{k, N_{k - 1} + 1}^\T \cdots \V{y}_{k, N_{k}}^\T]^\T$ and assume that new \ac{po} states are independent, i.e., $f(\overline{\V{y}}_{k}) = \prod_{n = N_{k - 1} + 1}^{N_k} f(\V{y}_{k, n})$.
We furthermore assume that the statistics of $\V{y}_{k, n}$ is based  
on a Poisson point process with mean $\mu_{\text{B}}$ and spatial \ac{pdf} $f_{\text{B}}(\V{x}_{k, n})$ \cite{Wil:J15}.
%Thus, new \ac{po} states $\V{y}_{k, n}$, $n \rmv\in\rmv \{N_{k - 1} + 1, \dots, N_{k}\}$ are independent, i.e., the \ac{pdf} is $\rd{f}(\overline{\V{y}}{}_{k}) = \prod_{n = N_{k - 1} + 1}^{N_k} \rd{f}(\V{y}_{k, n})$ with $\overline{\V{y}}{}_{k} = [\V{y}_{k, N_{k - 1} + 1}^\T \cdots \V{y}_{k, N_{k}}^\T]^\T$.
More precisely, we define 
%Let $\Set{X}_j$ be the subspace of $\Set{X}$ that consists of all kinematic states inside the area covered by measurement $\V{z}_{k, j}$ and let 
$f_{\text{B},j}(\V{x}_{k, n})$ being the birth pdf of new PO $n$ in data cell $j$ as equal to, up to a normalization constant, $f_{\text{B}}(\V{x}_{k, n})$ if $\V{x}_{k, n}$ is in cell $j$ and zero otherwise.
In order to define the birth probability, we first note that the expected number of new objects in cell $j$ is $\mu_{\text{B},j} = \mu_{\text{B}} \int_{\Set{X}_j} f_{\text{B}}(\V{x}_{k, n}) \ist \mathrm{d} \V{x}_{k, n}$ with $\Set{X}_j$ being the volume of cell $j$. By assuming that there is at most one new object in cell $j$, we obtain $p_{\text{B},j} = \mu_{\text{B},j} / (\mu_{\text{B},j} +1)$. Finally, our model for $f(\V{y}_{k, n})$ reads
%The \ac{pdf} of new \ac{po} $n  = N_{k - 1} + j$ can then be obtained\vspace{0mm} as 
\begin{equation}
    f(\V{y}_{k, n}) = f(\V{x}_{k, n}, r_{k, n}) = \begin{cases}
    (1 - p_{\text{B},j}) \ist f_{\text{D}}(\V{x}_{k, n}), & r_{k, n} = 0 \\
    p_{\text{B},j} \ist f_{\text{B},j}(\V{x}_{k, n}), & r_{k, n} = 1\ist.\\
    \end{cases} \nn \vspace{-2mm} 
\end{equation}
%with birth probability $p_{\text{B},j} = \mu_{\text{B},j} / (\mu_{\text{B},j} +1)$.

%up to a normalization constant for all 

%$\V{x}_{k, n} \rmv\in\rmv \Set{X}_j$ and equal to zero for all $\V{x}_{k, n} \rmv\in\rmv \Set{X} \backslash \Set{X}_j$. 
%Note that the expected number of new objects in $\Set{X}_j$ is also Poisson distributed with mean $\mu_{\text{B},j} = \mu_{\text{B}} \int_{\Set{X}_j} f_{\text{B}}(\V{x}_{k, n}) \ist \mathrm{d} \V{x}_{k, n}$.
%
%We perform the additional approximation by assuming that at time $k$, at most one new object appears in each $\Set{X}_j$. 
%
%The \ac{pdf} of new \acp{po} $n  = N_{k - 1} + j$ can then be obtained\vspace{-2mm} as 
%\begin{equation}
    %\rd{f}(\V{y}_{k, n}) = \rd{f}(\V{x}_{k, n}, r_{k, n}) = \begin{cases}
    %(1 - p_{\text{B},j}) \ist f_{\text{D}}(\V{x}_{k, n}), & r_{k, n} = 0 \\
    %p_{\text{B},j} \ist f_{\text{B},j}(\V{x}_{k, n}), & r_{k, n} = 1\\
    %\end{cases} \nn
%\end{equation}
%with birth probability\vspace{-4.0mm} $p_{\text{B},j} = \mu_{\text{B},j} / (\mu_{\text{B},j} +1)$.

\subsection{Object Declaration and State Estimation} 
\label{subsec:sys_model_decest}
\vspace{-.5mm}
For each time step $k$, our ultimate goal is (i)  to declare whether \ac{po} $n \rmv\in\rmv \{1, \dots, N_k\}$ exists and (ii) to estimate the state of existing \acp{po}. In our Bayesian setting, this necessitates the computation of the posterior distributions 
%the posterior existence probabilities 
$f(r_{k, n} = 1 | \V{z}_{1  : k})$ and 
%the posterior \acp{pdf} of the kinematic state 
$\V{x}_{k, n}$, i.e., $f(\V{x}_{k, n} | r_{k, n} = 1, \V{z}_{1 : k})$. 
In fact, \ac{po} is declared to exist if its existence probabilities $f(r_{k, n} = 1 | \V{z}_{1  : k})$ is larger than a chosen threshold $T_{\text{dec}}$ \cite[Ch. 2]{Poo:B94}. 
For existing \acp{po}, we perform \ac{mmse} state estimation according to $\hat{\V{x}}{}_{k, n} = \int \V{x}_{k, n} f(\V{x}_{k, n} | r_{k, n} = 1, \V{z}_{1 : k}) \mathrm{d}\V{x}_{k, n}$. Both $f(r_{k, n} = 1 | \V{z}_{1  : k})$ and $f(\V{x}_{k, n} | r_{k, n} = 1, \V{z}_{1 : k})$ can be obtained from the marginal posterior state \acp{pdf} $f(\V{y}_{k, n} | \V{z}_{1 : k})$. Thus, the remaining problem is to find an efficient method for calculating $f(\V{y}_{k, n} | \V{z}_{1 : k})$.

%-------------------------------------------------------------------------------------

%-------------------------------------------------------------------------------------
\section{Belief Propagation for \ac{tbd}}
\vspace{-.0mm}

\begin{figure}[!tbp]
    \centering
    \psfrag{da1}[c][c][0.85]{\raisebox{-2mm}{\hspace{.8mm}$\V{y}_{1}$}}
    \psfrag{daI}[c][c][0.85]{\raisebox{-2.5mm}{\hspace{.3mm}$\V{y}_{\underline{N}}$}}
    \psfrag{db1}[c][c][0.85]{\raisebox{-2mm}{\hspace{.2mm}$\V{y}_{\scriptscriptstyle \underline{N} + 1}$}}
    \psfrag{dbJ}[c][c][0.85]{\raisebox{1mm}{$\V{y}_{N}$}}
    \psfrag{q1}[c][c][0.85]{\raisebox{-1mm}{$f_{1}$}}
    \psfrag{qI}[c][c][0.85]{\raisebox{-3mm}{$f_{\underline{N}}$}}
    \psfrag{v1}[c][c][0.85]{\raisebox{-3.3mm}{\hspace{.15mm}$f_{\scriptscriptstyle \underline{N} + 1}$}}
    \psfrag{vJ}[c][c][0.85]{\raisebox{-1mm}{\hspace{.3mm}$f_{N}$}}
    \psfrag{g1}[c][c][0.85]{\raisebox{-1mm}{\hspace{.8mm}$p_{1}$}}
    \psfrag{gJ}[c][c][0.85]{\raisebox{-1.7mm}{\hspace{.5mm}$p_{J}$}}
    \psfrag{ma1}[c][c][0.85]{\color{blue}{$\alpha_{1}$}}
    \psfrag{maJ}[c][c][0.85]{\color{blue}{$\alpha_{N}$}}
    \psfrag{mb11}[l][l][0.85]{\color{blue}{$\beta_{1, 1}$}}
    \psfrag{mbJJ}[r][r][0.85]{\raisebox{-4mm}{\color{blue}{$\beta_{N, J}$}}}
    \psfrag{mk11}[r][r][0.85]{\color{blue}{\raisebox{1.5mm}{$\kappa_{1, 1}$}}}
    \psfrag{mkJJ}[l][l][0.85]{\color{blue}{$\kappa_{N, J}$}}
    \includegraphics[scale=0.9]{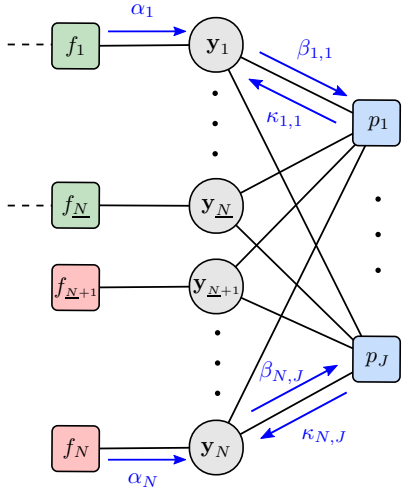}
    \caption{Factor graph representing a single \vspace{0.3mm} time step $k$ of $f(\V{y}_{0 : k} | \V{z}_{1 : k})$ in \eqref{eq:factorization}. We use the following short notation: $\underline{N} \rmv=\rmv N_{k - 1}$, \vspace{0.2mm} $N \rmv=\rmv N_k$, $\V{y}_n \rmv=\rmv \V{y}_{k,n}$, $p_{j} \rmv= f(\V{z}_{k, j} | \V{y}_k)$, $f_{n} = f(\V{y}_{k, n} | \V{y}_{k - 1, n})$ \vspace{0.4mm} for $n \rmv\in \{1, \dots, N_{k - 1}\}$, and $f_{n} = f(\V{y}_{k, n})$ for $n \rmv\in\rmv \{N_{k - 1} \rmv+\rmv 1, \dots, N_k\}$. Furthermore, 		
		$\alpha_{n} \rmv=\rmv \alpha_{k,n}(\V{y}_{k, n})$, $\beta_{n, j} \rmv= \beta_{k, n, j}^{(\ell)}(\V{y}_{k, n})$, and $\kappa_{n, j} \rmv= \kappa_{k, n, j}^{(\ell)}(\V{y}_{k, n})$.}
    % \caption{The factor graph at a single time step $k$ corresponding to the factorization \eqref{eq:factorization}. The following short notations are used: \( \underline{N} = N_{k - 1}, N = N_k, J = J_k\), \( \alpha_{n} = \alpha_{n, k}(\V{y}_{k, n})\), \(\beta_{n, j} = \beta_{k, n, j}^{(\ell)}(\V{y}_{k, n})\), \(\kappa_{n, j} = \kappa_{k, n, j}^{(\ell)}(\V{y}_{k, n})\), \(g_{j} = p(\V{z}_{k, j} | \V{y}_k)\), \(q_{n} = q(\V{y}_{k, n} | \V{y}_{k - 1, n})\) for \(n \in \{1, \dots, N_{k - 1}\}\), and \(q_{n} = q(\V{y}_{k, n})\) for \(n \in \{N_{k - 1} + 1, \dots, N_k\}\).}
    \label{fig:factor_graph}
   \vspace{-5mm}
\end{figure}

Based on the system model introduced in Section \ref{sec:sys_model}, common assumptions \cite{MeyKroWilLauHlaBraWin:J18}, and Bayes' rule, the joint posterior \ac{pdf} $f(\V{y}_{0 : k} | \V{z}_{1 : k})$ can be factorized as\vspace{-1mm}
\begin{align}
    \hspace{-0.5mm}f(\V{y}_{0 : k} | \V{z}_{1 : k}) &\propto \bigg( \prod_{n = 1}^{N_0} f(\V{y}_{0, n} ) \bigg) \prod_{k^\prime = 1}^{k} \bigg( \prod_{n = 1}^{N_{k^\prime - 1}} f(\V{y}_{k^\prime\, n} | \V{y}_{k^\prime - 1, n}) \bigg) \nn \\[1mm]
    & \times \bigg(  \prod_{n' \ist=\ist N_{k^\prime - 1} + 1}^{N_{k^\prime}} \rmv\rmv\rmv f(\V{y}_{k^\prime\rmv, n'}) \rrmv \bigg)  \prod_{j = 1}^{J} \ist p (\V{z}_{k^\prime, j} | \V{y}_{k^\prime}).    \label{eq:factorization} \\[-5mm]
    \nn
\end{align}
 %Often there are no \acp{po} at time $k = 0$, i.e., $N_0 = 0$. 
Given factorization \eqref{eq:factorization}, a factor graph \cite{KscFreLoe:01, Loe:04} representing the joint posterior \ac{pdf} $f(\V{y}_{0 : k} | \V{z}_{1 : k})$ can be constructed. Fig.~\ref{fig:factor_graph} shows a single time step of this graph.

\ac{bp} \cite{KscFreLoe:01,YedFreWei:05,KolFri:B09,Loe:04} performs local operations on factor graphs to compute representations of marginal posterior \acp{pdf}, called ``beliefs''. Since the factor graph in Fig.~\ref{fig:factor_graph} has loops, the beliefs $\tilde{f}(\V{y}_{k, n})$ provided by \ac{bp} are approximations of the true marginal posteriors $f(\V{y}_{k, n} | \V{z}_{1 : k})$, and there are many possible message passing orders \cite{KscFreLoe:01}. We only send \ac{bp} messages forward in time and perform iterative message passing at each time step individually \cite{MeyKroWilLauHlaBraWin:J18}. 
Next, we will present the specific \ac{bp} messages passed on the graph in Fig\vspace{-1.5mm}.~\ref{fig:factor_graph}. 
%-------------------------------------------------------------------------------------

%-------------------------------------------------------------------------------------

%-------------------------------------------------------------------------------------

%-------------------------------------------------------------------------------------

\subsection{Prediction and Birth Messages}
First, for each legacy PO $n \in \{1, \dots, N_{k-1}\}$, a prediction step is performed to compute the messages $\alpha_{k,n}(\V{y}_{k,n}) = \alpha_{n}(\V{x}_{k,n}, r_{k,n})$ that are passed from the factor nodes  ``$f(\V{y}_{k,n} | \V{y}_{k-1,n})$'' to the variable nodes ``$\V{y}_{k,n}$'', i.e.,
\begin{align}
 \alpha_{k,n}(\V{x}_{k,n}, 1) \rmv=\rmv \int \rmv\rmv p_{\mathrm{s}} \ist f(\V{x}_{k,n} | \V{x}_{k-1,n}) \tilde{f}(\V{x}_{k-1,n}, 1) \ist \mathrm{d} \V{x}_{k-1,n} \nn \\[-2mm]
  \label{eq:bp_alpha} \\[-7mm]
  \nn
\end{align}
and $\alpha_{k,n}(\V{x}_{k,n}, 0) \rmv=\rmv \alpha_{k,n}\ist f_{\text{D}}(\V{x}_{k,n})$. Here, $\alpha_{k,n}$ is the predicted probability of object non-existence \cite{MeyKroWilLauHlaBraWin:J18}. 
%determined by $\tilde{f}(\V{x}_{k - 1, n}, r_{k - 1, n})$ and $p_{\mathrm{s}}$. 
For new \acp{po} $n \rmv\in\rmv \{N_{k-1} \rmv+\rmv 1,\dots, N_k\}$, the messages from factor nodes ``$f(\V{y}_{k,n})$'' to variable nodes ``$\V{y}_{k,n}$'' are simply \cite{KscFreLoe:01} 
%the singleton factors themselves \cite{KscFreLoe:01}, i.e., 
$\alpha_{k,n}(\V{y}_{k,n}) = f(\V{y}_{k,n})$. 
In order to ease the notation, we omit the time index $k$ in the following and, e.g., simply write $\V{y}_{n}$ instead of $\V{y}_{k, n}$.
\vspace{-2mm}

\subsection{Iterative Message Passing and Belief Calculation}
\label{sec:iterativeMessagePassing}
We now perform iterative message passing between the variable nodes ``$\V{y}_{n}$'', $n \rmv\in\rmv \{1, \dots, N\}$, and the factor nodes ``$f(\V{z}_{j} | \V{y})$'',  $j \rmv\in\rmv \{1, \dots, J\}$. More precisely, at message passing iteration $\ell \rmv\in\rmv \{1, \dots, L\}$, 
%the following messages are computed for all \acp{po}, $n \in \{1, \dots, N\}$ and all measurements $j \in \{1, \dots, J\}$ in parallel. For $\ell \rmv>\rmv 1$, 
the messages $\beta_{n, j}^{(\ell)}(\V{y}_{n})$ are passed from the variable nodes ``$\V{y}_{n}$''  to the factor nodes ``$f(\V{z}_{j} | \V{y})$''.  For $\ell > 1$, these messages can be computed according to
\vspace{-1mm}
\begin{equation}
    \beta_{n, j}^{(\ell)}(\V{y}_{n}) = \frac{1}{C_{n,j}}\alpha_{n}(\V{y}_{n}) \prod_{\substack{j^\prime = \ist1 \\ j^\prime \ne\ist j}}^{J} \kappa_{n, j^\prime}^{(\ell - 1)}(\V{y}_{n};\V{z}_{j'})\ist, \label{eq:bp_beta}
		\vspace{0mm}
\end{equation}
and for $\ell \rmv=\rmv 1$, we set them to $\beta_{n, j}^{(1)}(\V{y}_{n}) = \alpha_{n}(\V{y}_{n})$. \vspace{-0.2mm}
Note that, $C_{n,j}$ is a normalization factor that ensures $\sum_{\V{y}_n} \rmv \beta^{(\ell)}_{n, j}(\V{y}_{n}) \rmv=\rmv 1$.
%, and for $\ell \rmv=\rmv 1$, we set them to $\beta_{n, j}^{(1)}(\V{y}_{n}) = \alpha_{n}(\V{y}_{n})$.

The messages $\kappa_{n, j}^{(\ell)}(\V{y}_{n})$ passed from the factor nodes ``$f(\V{z}_{j} | \V{y})$'' to the variable nodes ``$\V{y}$'' are obtained for $\ell \rmv\ \{1,\dots,L\}$ as
\vspace{-1mm}
\begin{equation}
    \kappa_{n, j}^{(\ell)}(\V{y}_{n};\V{z}_{j}  ) = \sum_{\V{y} \backslash \V{y}_{n}} f(\V{z}_{j} | \V{y}) \prod_{\substack{n^\prime =\ist 1 \\ n^\prime \ne\ist n}}^{N} \beta_{n^\prime\rmv, j}^{(\ell)}(\V{y}_{n^\prime}).\label{eq:bp_kappa}
		\vspace{-0.5mm}
\end{equation}
Here, $\sum_{\V{y} \backslash \V{y}_{n}}$ denotes 
\vspace{-0.4mm}
marginalization for all $\V{y}$ except $\V{y}_{n}$. This marginalization involves integration for continuous random vectors $\V{x}_n'$ and summation for binary random variables $r_n'$, $n' \in \{1,\dots,N\} \backslash \{n\}$. Note that our notation $\kappa_{n, j}^{(\ell)}(\V{y}_{n};\V{z}_{j})$ indicates that at this point the measurement $\V{z}_{j}$ is already observed and thus fixed. 
%When BP message passing is performed, $\V{z}_{j}$ is fixed and $\kappa_{n, j}^{(\ell)}(\V{y}_{n};\V{z}_{j})$ is only a function of $\V{y}_{n}$. However, we are still keeping $\V{z}_{j}$ in the notation of $\kappa_{n, j}^{(\ell)}(\V{y}_{n};\V{z}_{j}  )$ to motivate the scalable approximation of $\kappa_{n, j}^{(\ell)}(\V{y}_{n};\V{z}_{j}  )$ introduced in Section~\ref{subsec:bp_approx}.

After the last iteration $\ell = L$, the beliefs $\tilde{f}(\V{y}_{n})$ for all \acp{po} can be calculated as the normalized product of all incoming messages \vspace{0mm} according to
\begin{equation}
    \tilde{f}(\V{y}_{n}) = \frac{1}{C_{n}} \alpha_{n}(\V{y}_{n}) \prod_{j = 1}^{J} \kappa_{n, j}^{(L)}(\V{y}_{n};\V{z}_{j}). \label{eq:bp_belief}
\vspace{-1mm}
\end{equation}
Here, $C_{n}$ again ensures \vspace{.4mm} $\sum_{\V{y}_n} \tilde{f}(\V{y}_{n}) \rmv=\rmv 1$. 
The obtained beliefs can then be used for object declaration and state estimation as discussed in\vspace{-3.5mm} Section \ref{subsec:sys_model_decest}. 
 
\subsection{Approximate Computation of $\kappa_{n, j}^{(\ell)}(\V{y}_{n})$ and Complexity}
\label{subsec:bp_approx}

The computation of $\kappa_{k, n, j}^{(\ell)}(\V{y}_{k, n};\V{z}_{j})$ in \eqref{eq:bp_kappa}, relies on a high-dimensional marginalization whose complexity scales exponentially with the number of \acp{po} $N_k$. 
To improve this complexity scaling, we approximate $\kappa_{k, n, j}^{(\ell)}(\V{y}_{k, n};\V{z}_{j})$ as follows.
We interpret $\kappa_{n, j}^{(\ell)}(\V{y}_{n};\V{z}_{j})$ as the \ac{pdf} of $\V{z}_{j}$ and approximate it by a Gaussian \ac{pdf} via moment matching~\cite{DavGar:22}, i.e.,
%To reduce computational complexity and improve scalability, we make use of the fact that, for each $\V{y}_{n}$, $\kappa_{n, j}^{(\ell)}(\V{y}_{n};\V{z}_{j}  )$ can be interpreted as a \ac{pdf} of $\V{z}_{j}$ and we perform a Gaussian approximation of $\kappa_{n, j}^{(\ell)}(\V{y}_{n};\V{z}_{j})$ via moment matching  \cite{DavGar:22}. 
$\kappa_{n, j}^{(\ell)}(\V{y}_{n};\V{z}_{j}) \approx \tilde{\kappa}_{n, j}^{(\ell)}(\V{y}_{n};\V{z}_{j})$. Here, $\tilde{\kappa}_{n, j}^{(\ell)}(\V{y}_{n};\V{z}_{j}) =  \Set{N} \big( \V{z}_{j}; \V{\mu}^{(\ell)}_{\kappa, j}(\V{y}_{n}), \M{C}^{(\ell)}_{\kappa, j}(\V{y}_{n}) \big)$, where $\V{\mu}^{(\ell)}_{\kappa, j}(\V{y}_{n})$ is the matched mean and $\M{C}^{(\ell)}_{\kappa, j}(\V{y}_{n})$ is the matched covariance matrix, i.e., they are equal to the mean and covariance matrix of $\kappa_{n, j}^{(\ell)}(\V{y}_{n};\V{z}_{j})$. As derived in \cite{LiaKroMey:J23SM}, $\V{\mu}^{(\ell)}_{\kappa, j}(\V{y}_{n})$ and $\M{C}^{(\ell)}_{\kappa, j}(\V{y}_{n})$ \vspace{-.5mm} are given by 
\begin{align}
    \V{\mu}^{(\ell)}_{\kappa, j}(\V{y}_{n}) &= r_{n} \ist \V{\mu}_j(\V{x}_{n}) + \sum_{\substack{n^\prime = 1 \\ n^\prime \ne n}}^{N} \V{\mu}^{(\ell)}_{n^\prime\rmv, j} \label{eq:bp_kappa_mean} \\[.5mm]
    \M{C}^{(\ell)}_{\kappa, j}(\V{y}_{n}) &= r_{n }\M{C}_j(\V{x}_{n}) + \M{C}_{\V{\epsilon}}  \nn \\[0mm]
    &\hspace{5mm}+ \sum_{\substack{n^\prime = 1 \\ n^\prime \ne n}}^{N} \rmv\rmv\rmv \Big( \M{R}^{(\ell)}_{n^\prime, j} -  \V{\mu}^{(\ell)}_{n^\prime j} \ist \V{\mu}^{(\ell)  \T}_{n^\prime\rmv, j} \Big). \label{eq:bp_kappa_cov}
\end{align}
Here, we have\vspace{.5mm} introduced $\V{\mu}^{(\ell)}_{n, j} \triangleq \text{E}_{n, j}^{(\ell)} \big( \V{\mu}_j(\V{x}_{n}) \big)$ and $\M{R}^{(\ell)}_{n, j} \triangleq \text{E}_{n, j}^{(\ell)} \big( \M{C}_j(\V{x}_{n}) + \V{\mu}_j(\V{x}_{n}) \V{\mu}_j^\T(\V{x}_{n}) \big)$ which use the notation  $\text{E}_{n, j}^{(\ell)}( \cdot ) \triangleq \int \cdot \ist\ist \beta^{(\ell)}_{n, j}(\V{x}_{n}, 1) \hspace{1mm} \mathrm{d} \V{x}_{n}$.

In summary, our \ac{bp}-based \ac{tbd} algorithm consists of executing the following steps for each time step $k$: First, we perform state prediction according to \eqref{eq:bp_alpha} for $n \rmv\in\rmv \{1,\ldots,N_k\}$. We then run the iterative message passing scheme by iteratively computing \eqref{eq:bp_beta} and \eqref{eq:bp_kappa_mean}--\eqref{eq:bp_kappa_cov} for $\ell \rmv\in\rmv \{1,\ldots,L\}$. Finally, after computing the \ac{po} beliefs according to \eqref{eq:bp_belief}, object declaration and state estimation are performed as described in Section~\ref{subsec:sys_model_decest}.
An inspection of all these computations shows that the operations with the highest complexity are the summations in \eqref{eq:bp_kappa_mean} and \eqref{eq:bp_kappa_cov}, whose complexity scales according to $\Set{O}(N J)$. We can therefore conclude that the computational complexity of the proposed algorithm also exhibits this scaling behavior\vspace{-3mm}.

%Computing $\sum_{n = 1}^{N} \V{\mu}^{(\ell)}_{n, j}$ and $\sum_{n = 1}^{N} \big( \M{R}^{(\ell)}_{n, j} - \V{\mu}^{(\ell)}_{n, j} \ist \V{\mu}^{(\ell) \T}_{n, j} \big)$ for each $J$ yields a complexity that scales as $\Set{O}(N J)$. After these precomputations, evaluation of \eqref{eq:bp_kappa_mean} and \eqref{eq:bp_kappa_cov} has a constant complexity. The total computation complexity related to computing the $N J$ messages $\tilde{\kappa}_{n, j}^{(\ell)}(\V{y}_{n}; \V{z}_{j})$ is thus $\Set{O}(N J)$. It can easily be verified that the overall complexity that also involves computation of messages and beliefs  \eqref{eq:bp_alpha}, \eqref{eq:bp_beta}, and \eqref{eq:bp_belief}, still only scales as\vspace{-2mm} $\Set{O}(N J)$.

%-------------------------------------------------------------------------------------

%-------------------------------------------------------------------------------------
% \section{Particle-Based Implementation}

%-------------------------------------------------------------------------------------

%-------------------------------------------------------------------------------------
\section{Numerical Results}
\label{sec:exp}
\vspace{-.5mm}

%\subsection{Experimental Setup} \label{subsec:exp_setup}
We consider a two-dimensional (2D) simulation scenario with a \ac{roi} of $[0\text{m}, 32\text{m}] \times [0\text{m}, 32\text{m}]$.
We simulated five objects and 50 time steps.
%, where the time interval between two consecutive time steps is $1$s.
The object states are modeled by random vectors $\V{x}_{k, n} = [\V{p}_{k, n}^\T \hspace{1mm} \V{v}_{k, n}^\T \hspace{1mm} \gamma_{k, n}]^\T$ consisting of 2D position $\V{p}_{k, n}$, 2D velocity $\V{v}_{k, n}$, and the object's intensity $\gamma_{k, n}$. 
The objects appear at time steps $k \in \{1, 5, 10, 15, 20\}$ at positions randomly chosen in the region $[8\text{m}, 24\text{m}] \times [8\text{m}, 24\text{m}]$. 
The object's initial velocity is drawn from $\Set{N}(\V{v}_{\cdot, n}; \V{0}, 10^{-2}\M{I}_2)$ and the object's initial intensity is $\gamma_0$.  
The object's position $\V{p}_{k, n}$ and velocity $\V{v}_{k, n}$ evolve according to a constant velocity model \cite[Ch. 4]{ShaKirLi:B02} with \ac{iid} zero-mean Gaussian noise with variance $10^{-3}$ \cite[Ch. 4]{ShaKirLi:B02}. The object's intensity $\gamma_{k, n}$ evolves according to a random walk model with \ac{iid} zero-mean Gaussian noise with variance $10^{-4}$. 
The objects disappear at $k \in \{31, 36, 41, 46, +\infty\}$ or when they leave the \ac{roi}.

%A constant velocity model is used as the state-transition model for the motion-related variables $\V{p}_{k, n}$ and $\V{v}_{k, n}$ with Gaussian driving noise $\Set{N}(\V{0}, 10^{-3} \M{I})$ \cite[Ch. 4]{ShaKirLi:B02}, and a random walk model is used for the intensity variable $\gamma_{k, n}$ with Gaussian driving noise $\Set{N}(0, 10^{-4})$. The 5 objects are randomly initialized over the region $[8, 24]\text{m} \times [8, 24]\text{m}$ at time steps $k \in \{1, 5, 10, 15, 20\}$, and disappear at $k \in \{30, 35, 40, 45, 50\}$ or when they are out of the \ac{roi}. 

%The objects’ initial velocity is drawn from $\Set{N}(\V{0}, 10^{-2}\M{I})$ and the initial intensity is $\gamma_0$. The survival probability is $p_{\mathrm{s}} = 0.999$.

The measurement $\V{z}_k$ is an image of $32 \times 32$ pixels or bins, i.e., $J = 1024$. Each bin has a square size with $1$m length, covering the total \ac{roi}.
 %and each pixel in the image is a square with 1m width. 
Pixel $\V{z}_{k,j}$, $j \rmv=\rmv 1,\ldots,J$ is represented by the 2D vector with center position $\V{p}_{j}^{z}$.
We use the measurement model defined by \eqref{eq:meas_model} and set the mean and the covariance of the Gaussian random vector $\V{h}_{j,k, n}$
%, which describes $p(\V{h}_{k, j, n} | \V{x}_{k, n})$, 
to $\V{\mu}_j(\V{x}_{k, n}) = \V{0}$ and \vspace{-2mm}
%The conditional \ac{pdf} of the contribution $\V{h}_{k, j, n}$ in \eqref{eq:meas_model} is Gaussian and we set $\V{\mu}_j(\V{x}_{k, n}) = \V{0}$ and 
\begin{equation}
    \M{C}_j(\V{x}_{k, n}) = \frac{\gamma_{k, n}}{2 \pi \sigma_S^2} \exp \Big( -\frac{\Vert \V{p}_{k, n} - \V{p}_j^{z} \Vert^2}{2 \sigma_S^2} \Big) \M{I}_2,
\end{equation}
respectively. 
Note that the variance $\sigma_S^2$ defines the shape of $f\big(\V{h}_{j,k, n})\big)$ and thus the number of pixels illuminated by object $n$.   
Furthermore, the covariance of the noise vector $\V{\epsilon}_{k, j}$ in \eqref{eq:meas_model} is set to $\M{C}_{\V{\epsilon}} = \sigma_{\epsilon}^2\M{I}_2$ with $\sigma_{\epsilon}^2 = 1$. 
These settings lead to a measurement process in which the contribution of object $n$ on pixel $j$ is large if the intensity $\gamma_{k, n}$ of object $n$ is large and the position $\V{p}_{k, n}$ of object $n$ is close to pixel $j$.

%The higher the intensity value $\gamma_{k, n}$ and the closer the \ac{po} to a pixel, the more likely it has a large contribution to the pixel. The covariance of the noise is set to $\M{C}_{\V{\epsilon}} = \sigma_{\epsilon}^2\M{I}$ with $\sigma_{\epsilon}^2 = 1$. Gaussian approximation of messages $\kappa_{k, n, j}^{(\ell)}(\V{y}_{k, n})$ is applied as discussed in Section \ref{subsec:bp_approx}.

We employ a particle implementation of our proposed \ac{bp}-based \ac{tbd} algorithm denoted by TBD-BP \cite{IhlMca:09,MeyBraWilHla:J17}.
%, that is closely related to the particle implementations in \cite{IhlMca:09,MeyBraWilHla:J17}. 
%More precisely, the integrations in \eqref{eq:exp1} and \eqref{eq:exp2} are computed using Monte Carlo Integration \cite{DouFreGor:01}, and the messages and beliefs in \eqref{eq:bp_beta} and \eqref{eq:bp_belief} are calculated by means of importance sampling \cite{DouFreGor:01}.
The spatial \ac{pdf} of each \ac{po} state's belief is represented by 3000 particles. 
The generation of new \ac{po}s is based on the measurement  $\V{z}_k$, in particular on $\Vert \V{z}_{k, j} \Vert$.
To keep the computational complexity low, we initialize a new \ac{po} $n \in \mathcal{N}_{n} \subseteq \{N_{k - 1} + 1,\dots, N_k\}$ only for those pixels whose intensity $\Vert \V{z}_{k, j} \Vert$ is larger than the predefined threshold $1.5\sqrt{ \gamma_0 \ist / \ist (2 \pi \sigma_S^2) + \sigma_{\epsilon}^2}$.    
The spatial \ac{pdf} of new \ac{po} $n \in \mathcal{N}_{n}$ is modeled by $f_{\text{B},j}(\V{x}_{k, n}) = f_j(\V{p}_{k, n}) f(\V{v}_{k, n}) f(\gamma_{k, n})$.
Here, $f_j(\V{p}_{k, n})$ is uniform over the area of pixel $j$, $f(\V{v}_{k, n})$ is $\Set{N}(\V{v}_{k, n}; \V{0}, 10^{-2}\ist\M{I}_2)$, and $f(\gamma_{k, n})$ is uniform from $0$ to $\gamma_{\text{max}} = 2\ist\gamma_0$.
The birth probability  of each new \ac{po} is set to $p_{\text{B}, j} = 10^{-5}$. 
An object is declared to exist if its existence probability is larger than $T_{\text{dec}} = 0.5$. 
%On the other hand, since the number of \acp{po} grows linearly with time $k$, we prune all \acp{po} whose existence probability is below $T_{\text{pru}} = 10^{-3}$.
We prune \acp{po} whose existence probability is below $T_{\text{pru}} = 10^{-3}$ and set the survival probability to $p_{\mathrm{s}} = 0.999$.

\begin{figure}[!tbp]
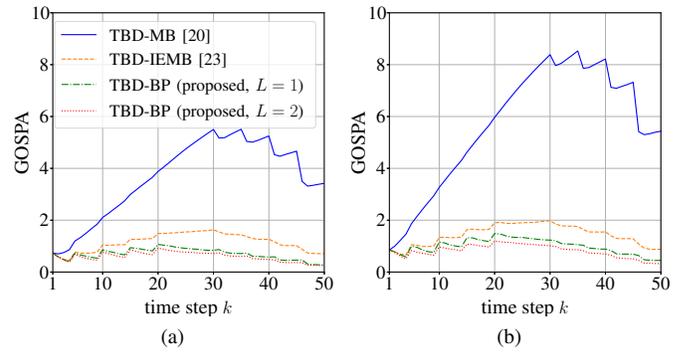

    \centering
    \subfloat[]{\resizebox{0.48\linewidth}{!}{\input{Figs/GOSPA_s0.5_p3e2_linestyle.pgf}}} \hspace{1mm}
    \subfloat[]{\resizebox{0.48\linewidth}{!}{\input{Figs/GOSPA_s1_p3e2_linestyle.pgf}}}
    \caption{GOSPA error of TBD-MB, TBD-IEMB and the proposed TBD-BP versus time $k$ for $\gamma_0 = 60$, (a) $\sigma_S^2 = 0.5$ and (b) $\sigma_S^2 = 1$.}
    \label{fig:gospa_baseline}
\end{figure}

%\subsection{Performance Evaluation} 
%\label{subsec:exp_result}
To evaluate the performance of our proposed algorithm, we compute the Euclidean distance based \ac{gospa} metric \cite{RahGarSve:17} averaged over 400 simulation runs, with cutoff parameter $c = 1$, order $p = 2$, and $\alpha \rmv=\rmv 2$.
%The lower the \ac{gospa}, the better the tracking performance. 
In our first experiment, we compare the proposed TBD-BP with a particle implementation of the \ac{mb} filter in \cite{VoVoPhaSut:10}, referred to as TBD-MB, and the \ac{iemb} filter in \cite{DavGar:22}. The TBD-IEMB models spatial distributions by Gaussian \acp{pdf}.
% In our first experiment, we compare the proposed TBD-BP with a particle implementation of the \ac{tbd} \ac{mb} filter in \cite{VoVoPhaSut:10}, denoted by TBD-MB. 
%We use \cite{VoVoPhaSut:10} as our reference method, denoted by ``TBD-MB''. The setup of TBD-MB is exactly the same as the proposed TBD-BP as discussed in Section \ref{subsec:exp_setup}. 
The \ac{gospa} results are displayed in Fig. \ref{fig:gospa_baseline}.
 %we present the \ac{gospa} results of TBD-MB and our proposed method TBD-BP with different numbers of message passing iterations $L$.
As the figure shows, TBD-BP with $L = 2$ message passing iterations performs slightly better than TBD-BP with $L = 1$, followed by TBD-IEMB, and significantly better than TBD-MB.
%Recap that $L$ is the number of message passing iterations (cf. Section \ref{sec:iterativeMessagePassing}).
%The huge performance gain of the proposed TBD-BP method compared to TBD-MB can be attributed to the fact that the iterative message passing scheme of TBD-BP systematically takes into account the interaction between \acp{po}, while TBD-MB tracks objects independently.
The lower performance of TBD-MB is due to the fact that it does not model the interaction of objects and tracks them independently.   
%The huge performance gain of the proposed TBD-BP method compared to TBD-MB can be attributed to the fact that the iterative message passing scheme of TBD-BP systematically takes into account the interaction between \acp{po}, while TBD-MB tracks objects independently. 
Furthermore, the track initialization scheme of TBD-MB leads to a high number of false tracks resulting in an almost linear increase in \ac{gospa} for $k < 30$. 
%\rd{The slightly lower performance of TBD-IEMB compared to TBD-BP also considers can be explained by the fact that, although it models the interaction between objects, it models object states as Gaussian assumption on object state posteriors. }

%TBD-MB keeps initializing new \acp{po} even in places where there are \acp{po} from previous time steps, leading to many false tracks and consequently to a linear increase in GOSPA for the first 30 time steps.

%This is because TBD-BP considers the influence among \acp{po} via iterative message passing (cf. \eqref{eq:bp_kappa} and \eqref{eq:bp_beta}), while TBD-MB does not take this into account. 

%As a result, TBD-MB keeps initializing new \acp{po} even in places where there are \acp{po} from previous time steps, leading to many false tracks and consequently to a linear increase in GOSPA for the first 30 time steps. 
%We also notice that TBD-BP with $L = 2$ has a better performance than the one with $L = 1$. This can be attributed to the additional message passing iteration that helps provide a more accurate approximation of the true marginal posterior distributions.

\begin{figure}[!tbp]
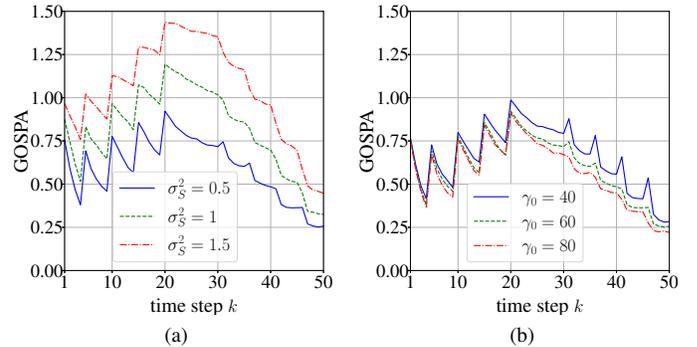

    \centering
    \vspace{-5mm}
    \mbox{\hspace{0mm}\subfloat[]{\resizebox{0.48\linewidth}{!}{\input{Figs/GOSPA_sigma_p3e2_linestyle.pgf}} \label{subfig:sigma}} \hspace{1.0mm}
    \subfloat[]{\resizebox{0.48\linewidth}{!}{\input{Figs/GOSPA_gamma_p3e2_linestyle.pgf}} \label{subfig:gamma}}}
    \caption{GOSPA metric the proposed TBD-BP ($L = 2$) with (a) $\gamma_0 = 60$ and different $\sigma_{S}^2 $; (b) $\sigma_S^2 = 0.5$ and different $\gamma_0$.}
    \label{fig:gospa_gamma_sigma}
    \vspace{-4.5mm}
\end{figure}

In the second experiment, we investigated the GOSPA performance of TBD-BP for different intensity values $\gamma_0$ and variances $\sigma_S^2$. 
As Fig. \ref{subfig:sigma} shows, the \ac{gospa} error increases for larger $\sigma_S^2$.
This is due to the fact that for larger $\sigma_S^2$, an object illuminates a larger number of neighboring pixels, which in turn leads to more false tracks caused by the now larger number of pixels of higher intensity.  
Fig. \ref{subfig:gamma} additionally shows that the \ac{gospa} error decreases as $\gamma_0$ increases\vspace{-3mm}. 
%This is expected as larger $\gamma_0$ leads to higher \ac{snr}. 

%In addition, we conducted experiments on TBD-BP with different $\gamma_0$ and $\sigma_S^2$. From Fig. \ref{subfig:sigma} we can see that the \ac{gospa} error generally increases as $\sigma_S^2$ increases, due to the fact that the contribution of a \ac{po} spreads out to neighboring pixels when $\sigma_S^2$ is large. In this case, there are more measurements with high intensity $\Vert \V{z}_{k, j}\Vert$ and the algorithm tends to initialize more false tracks. Fig. \ref{subfig:gamma} shows that the tracking performance increases as $\gamma_0$ increases. This is expected as larger $\gamma_0$ leads to higher \ac{snr}. 

%-------------------------------------------------------------------------------------
% GOSPA
%-------------------------------------------------------------------------------------

%-------------------------------------------------------------------------------------

%-------------------------------------------------------------------------------------
\vspace{0mm}
\section{Conclusion}
\vspace{-1mm}

In this letter, we propose a \ac{bp} method for tracking an unknown number of low-observable objects based on the \ac{tbd} approach. We introduced a new object birth model and a new measurement model that allows interacting objects to contribute to more than one data cell. To reduce computational complexity and improve scalability, certain \ac{bp} messages are approximated by Gaussian distributions. Experiments conducted on image data show that the proposed TBD-BP method outperforms two state-of-the-art \ac{tbd} \ac{mb} filtering methods \cite{VoVoPhaSut:10,DavGar:22}. However, the proposed method is limited to measurements that are independent conditioned on the object states. An interesting possibility for future research is the design of a more general measurement model and an application to real data.
Other promising directions for future research are the development of \ac{tbd} approaches for extended object tracking \cite{MeyWil:J21} or based on particle flow \cite{DuaHua:07,JanMeySny:J23,LiyDau:J23,ZhaMey:J23} and an extension to hybrid model-based and data-driven \ac{tbd} \cite{LiaMey:J23}.

\bibliographystyle{IEEEtran}
\bibliography{IEEEabrv,StringDefinitions,Books,Papers,ref,refBooks}

\newpage

\end{document}